\documentclass[conference]{IEEEtran}
\usepackage{amsmath,amssymb,amsthm,bm}
\usepackage{booktabs}
\usepackage{graphicx}
\usepackage{cite}
\usepackage[hidelinks]{hyperref}
\graphicspath{{images/}}
\newtheorem{proposition}{Proposition}

\title{Lost Opportunity Costs Under Ramp Stress:\\
A Comparison of Ramp-Product Dispatch and Look-Ahead Economic Dispatch}
\author{\IEEEauthorblockN{Aidan Looney, Qian Zhang, and Le Xie}
\IEEEauthorblockA{
Harvard John A. Paulson School of Engineering and Applied Sciences\\
Allston, MA, USA\\
Email: \{aidanlooney, qianzhang\}@g.harvard.edu, xie@seas.harvard.edu
}}

\begin{document}
\maketitle

\begin{abstract}
This paper studies whether ramp-product settlement compensates generators that absorb intertemporal ramp scarcity as effectively as look-ahead economic dispatch. We evaluate this question by comparing the lost opportunity cost (LOC) induced by each approach. Using rolling-horizon simulations on a 10-generator system and a modified RTS-GMLC system, we compare ramp-product settlement (RP-LMP), look-ahead settlement (LA-LMP), and temporal locational marginal pricing (TLMP). In the featured deterministic ramp stressed cases, aggregate LOC is higher under RP-LMP than under LA-LMP. The main contribution is a generator-level critical assessment for ramp-product settlement: we identify which units are left with uncompensated intertemporal opportunity cost under RP-LMP and how that burden changes under LA-LMP. RP-LMP LOC is concentrated on units with repeated ramp-binding exposure, while LA-LMP mainly relieves those same units and leaves smaller residual LOC on other units. Multi-day tests preserve this ordering under perfect foresight in the larger test system, but show that it need not hold under forecast error.
\end{abstract}


\section{Introduction}
As more renewable generation is integrated into electricity grids, net load (total load minus renewable generation) becomes increasingly variable. Generator ramp limits create an intertemporal coupling in real-time dispatch: a generator's current output affects its feasible output in later intervals. When these constraints bind, current-interval energy prices may fail to compensate generators for the flexibility they forgo by following dispatch, leading to incentive misalignment and lost opportunity cost (LOC).

Independent System Operators (ISOs) address system-wide ramping in two broad ways. One augments single-interval dispatch with ramping products co-optimized with energy (RP), preserving a largely myopic market structure while enforcing forward-looking feasibility. The other solves a multi-interval look-ahead economic dispatch (LAED), explicitly modeling the dispatch trajectory over a finite horizon. While both approaches incorporate intertemporal constraints into dispatch, generator settlement may still fail to price those intertemporal outcomes properly, even when a separate ramp product is cleared \cite{Zhang2026_PSCC}.

These approaches connect to three strands of literature. First, flexible ramping products and related designs have been studied as practical tools for managing renewable variability and short-term uncertainty while retaining single-interval pricing frameworks \cite{ela2016,Wang2017_review,CWang2017,wu2016,Chen2017,Chen2023}. Second, multi-interval dispatch formulations explicitly capture intertemporal feasibility by optimizing over a look-ahead horizon, but introduce more complex economic interpretations of prices due to intertemporal constraints \cite{Hua2019,Zhao2020,Scott2019,mickey2015mirtm}. Third, several pricing methods have been proposed to address intertemporal pricing, such as price-preserving pricing (PMP), constraint-preserving pricing (CMP), multi-settlement LMP (MLMP), and temporal locational marginal pricing (TLMP)\cite{hogan2020,Hua2019,Zhao2020,Guo2021}.

The key problem is therefore not only whether a ramp-product market preserves short-horizon feasibility, but whether it compensates the generators that actually absorb intertemporal scarcity as effectively as a look-ahead dispatch. We study this question using ex-post LOC, where each generator evaluates self-dispatch against the realized price trajectory \cite{Cho2023}, in a common rolling-horizon comparison of RP-LMP, LA-LMP, and TLMP.

We ask two central questions: which generators experience the highest LOC, and how does that burden change when dispatch moves from RP-LMP to LA-LMP? Unlike prior comparative studies focused mainly on operational reliability under these two dispatch methods \cite{Zhang2026_PSCC}, this paper contributes a pricing and generator-level incidence analysis. RP-LMP and LA-LMP are therefore evaluated as integrated dispatch--settlement designs; their LOC difference is not interpreted as the isolated causal effect of either look-ahead dispatch or pricing. The key innovation is to characterize how the two designs assign uncompensated ramp-scarcity costs across individual generators, rather than only comparing feasibility or total system cost. A product can procure ramp capability and still leave the opportunity cost concentrated on a small set of generators if the settlement does not price the intertemporal value of their dispatch trajectory. The contribution is threefold. First, we compare RP-LMP, LA-LMP, and TLMP in one rolling-horizon simulation framework and use ex-post LOC as a common diagnostic of uncompensated intertemporal value. Second, we decompose aggregate LOC into dispatch inefficiency and a price-support gap, clarifying why look-ahead dispatch can reduce but not eliminate LOC under uniform energy settlement. Third, we identify generator-level exposure features, especially binding frequency scaled by flexibility, that explain which units bear the RP-LMP burden and which receive LA-LMP relief. We test this incidence pattern on both the 10-generator system and a modified RTS-GMLC system.

\section{Dispatch Formulations and Pricing Rules}
\label{sec:models}
Consider a single-bus real-time dispatch problem. Let $\mathcal{G}=\{1,\dots,G\}$ denote the set of generators indexed by $g$ and let $\mathcal{W}=\{0,1,\dots,W\}$ indexed by $\tau$ denote the current and advisory intervals in a rolling-horizon formulation. Demand at time $t+\tau$ is denoted by $d_{t+\tau}$, generator output by $p_{g,t+\tau}$ with $\tau=0$ for the ramp-product formulation (denoted just $p_g$), and load shedding, $s_{t+\tau}\ge 0$. Each generator $g$ has linear marginal cost $c_g$, capacity bounds $\underline P_g=0,\overline P_g$, and upward and downward ramp limits $\overline{R}_g,\underline{R}_g$ per 5-minute dispatch interval. In the single-interval RP formulation, the previous realized output enters as data and is denoted by $p_{g,t-1}$. The value of lost load is denoted by $\rho$.

\subsection{Single-Interval Dispatch with Ramp Products}
The first dispatch method is a single-interval economic dispatch with co-optimized 10-minute ramp-capability products. The current dispatch decision is $p_g$, while $r^u_g$ and $r^d_g$ denote awarded upward and downward ramp-capability products. We model a simplified market-wide ramp-capability requirement, rather than a full ISO reserve-demand-curve implementation:
\begin{align}
\min_{p,r^u,r^d,s,z^u,z^d} \quad
& \sum_{g\in\mathcal{G}} c_g p_g + \rho s_t + \kappa^u z_t^u + \kappa^d z_t^d \label{eq:rp_obj}\\
\text{s.t.}\quad
& \sum_{g\in\mathcal{G}} p_g = d_t - s_t \label{eq:rp_balance}\\
& -\underline R_g \le p_g - p_{g,t-1} \le \overline R_g, \qquad \forall g \label{eq:rp_prev}\\
& 0 \le r^u_g \le K \overline R_g, \qquad \forall g \label{eq:rp_ru}\\
& 0 \le r^d_g \le K \underline R_g, \qquad \forall g \label{eq:rp_rd}\\
& p_g + r^u_g \le \overline P_g, \qquad \forall g \label{eq:rp_head}\\
& p_g - r^d_g \ge \underline P_g, \qquad \forall g \label{eq:rp_foot}\\
& \sum_{g\in\mathcal{G}} r^u_g + z_t^u \ge R_t^{u,\mathrm{req}} \label{eq:rp_sysu}\\
& \sum_{g\in\mathcal{G}} r^d_g + z_t^d \ge R_t^{d,\mathrm{req}} \label{eq:rp_sysd}\\
& s_t, z_t^u, z_t^d \ge 0, \qquad
  p_g, r^u_g, r^d_g \ge 0,\ \forall g. \label{eq:rp_shed}
\end{align}
Here $K=2$, representing two 5-minute intervals and therefore a 10-minute ramping product. The market-wide product requirements are constructed as
\begin{align}
R_t^{u,\mathrm{req}} = V_t^u + U_t^u, \qquad
R_t^{d,\mathrm{req}} = V_t^d + U_t^d, \label{eq:rp_req}
\end{align}
where $V_t^u,V_t^d$ are forecasted net-load-change components and $U_t^u,U_t^d$ are uncertainty adders. Following MISO \cite{MISO2026URC}, $U_t^u$ proxies its upward uncertainty adder, set either in MW or as $\beta\bar d$; values are not calibrated MISO requirements, and $U_t^d=0$. In the implementation used for the reported results,
\begin{align}
V_t^u = \max\{0, \widehat d_{t+K|t}-d_t\}, \\
V_t^d = \max\{0, d_t-\widehat d_{t+K|t}\} \label{eq:rp_var}
\end{align}
where $\widehat d_{t+K|t}$ is the 10-minute-ahead net-load forecast. The shortage costs $\kappa^u,\kappa^d$ are a single-segment scarcity proxy with associated slack variables $z_t^u,z_t^d$; in all reported simulations, $\kappa^u=\kappa^d=\$65/\text{MWh}$, anchored to the spinning-reserve violation cost in \cite{Chen2023}.

This RP formulation preserves the economic distinction between a single-interval capability market and multi-interval dispatch, but it is not a reproduction of an ISO implementation: it omits network and zonal requirements, multi-segment demand curves, eligibility and deployment rules, and other reserve products. Accordingly, the headline RP-LMP/LA-LMP difference compares complete dispatch-and-settlement bundles, not a pure pricing or horizon effect; matched ablations separate these channels in Section~\ref{sec:results}.

The generator settlement considered for this formulation pays current-interval energy at the balance price and awarded ramp capability at the system-wide ramp-product prices:
\begin{align}
\Pi_g^{RP}=\sum_t \left(\lambda_t^{RP} p_{g,t} + \nu_t^{u} r^u_{g,t} + \nu_t^{d} r^d_{g,t} - c_g p_{g,t}\right)\Delta t, \label{eq:rp_lmp}
\end{align}
where $\lambda_t$, $\nu_t^{u}$, and $\nu_t^{d}$ are the shadow prices on \eqref{eq:rp_balance}, \eqref{eq:rp_sysu} and \eqref{eq:rp_sysd} respectively. This price settlement models the co-optimization of the energy dispatch and ramp product market.

\subsection{Look-Ahead Economic Dispatch}
The look-ahead dispatch optimizes the full trajectory over the rolling horizon with re-optimization as the window rolls:
\begin{align}
\min_{p_{g,t+\tau},s_{t+\tau}} \quad
& \sum_{\tau\in\mathcal{W}}\left(\sum_{g\in\mathcal{G}} c_g p_{g,t+\tau} + \rho s_{t+\tau}\right) \label{eq:la_obj}\\
\text{s.t.}\quad
& \sum_{g\in\mathcal{G}} p_{g,t+\tau} = d_{t+\tau} - s_{t+\tau}, \qquad \forall \tau \label{eq:la_balance}\\
& \underline P_g \le p_{g,t+\tau} \le \overline P_g, \qquad \forall g,\tau \label{eq:la_cap}\\
& p_{g,t+\tau} - p_{g,t+\tau-1} \le \overline R_g, \qquad \forall g,\tau \label{eq:la_ru}\\
& p_{g,t+\tau-1} - p_{g,t+\tau} \le \underline R_g, \qquad \forall g,\tau \label{eq:la_rd}\\
& p_{g,t+\tau} \geq 0 \ \forall g, \tau\qquad s_{t+\tau} \ge 0, \  \forall \tau. \label{eq:la_shed}
\end{align}
The realized dispatch in the rolling simulation is the current-period component $p_{g,t}^{\star}$; the optimization is then advanced one interval and resolved with updated initial conditions.

We study two settlements for LAED.

\paragraph{LA-LMP}
The first is the current-period LMP in a single-bus setting,
\begin{align}
\pi_t^{LA} = \lambda_t^{LA}, \label{eq:la_lmp}
\end{align}
where $\lambda_t^{LA}$ is the shadow price of the current-period balance equation in \eqref{eq:la_balance}.

\paragraph{TLMP}
The second is TLMP. In the single-bus setting, the generator-specific TLMP at time $t$ can be written as
\begin{align}
\pi_{g,t}^{TLMP} = \lambda_t^{LA} + \left(\overline\mu_{g,t+1}-\underline\mu_{g,t+1}\right) - \left(\overline\mu_{g,t}-\underline\mu_{g,t}\right), \label{eq:tlmp}
\end{align}
where $\overline\mu_{g,t}$ and $\underline\mu_{g,t}$ are the dual variables associated with the ramping constraints \eqref{eq:la_ru} and \eqref{eq:la_rd}. TLMP is discriminatory and internalizes generator-specific ramping shadow values such that under convexity, it yields zero LOC \cite{Guo2021}.

\section{Lost Opportunity Cost and Generator-Level Explanatory Features}
\label{sec:loc}
We evaluate each settlement rule by comparing realized generator profit under ISO dispatch to profit under self-dispatch at the same price sequence. This ex-post metric is not a behavioral bidding model; it identifies which units are left with uncompensated intertemporal value after following dispatch.

\subsection{Generator Profit Maximization Problem}
For the energy-only settlements, each generator faces a realized settlement sequence $\pi_g = \{\pi_{g,t}\}_{t=1}^T$. Under LA-LMP this sequence is uniform across units, with $\pi_{g,t}^{LA}\equiv \lambda_t^{LA}$, while under TLMP it is generator specific as in \eqref{eq:tlmp}. The generator best-response formulation is: 
\begin{align}
Q_g(\pi_g) := \max_{q_{g,t}} \quad
& \sum_{t=1}^T \left( \pi_{g,t} - c_g \right) q_{g,t}\Delta t \label{eq:gen_br_obj}\\
\text{s.t.}\quad
& \underline P_g \le q_{g,t} \le \overline P_g, \qquad \forall t \label{eq:gen_cap}\\
& -\underline R_g \le q_{g,t} - q_{g,t-1} \le \overline R_g, \qquad \forall t \label{eq:gen_ramp}
\end{align}
with initial condition $q_{g,0}$ given.

For RP-LMP, the generator also responds to the ramp-product prices.
The corresponding self-scheduling problem is
\begin{align}
Q_g^{RP}:={}&
\max_{\substack{q_{g,t},\\ u_{g,t},d_{g,t}}}
\sum_{t=1}^T
\left[
(\lambda_t^{RP}-c_g)q_{g,t}
+\nu_t^{u}u_{g,t}
+\nu_t^{d}d_{g,t}
\right]\Delta t
\label{eq:gen_br_rp_obj}\\
\text{s.t.}\quad
&-\underline R_g
\le q_{g,t}-q_{g,t-1}
\le \overline R_g,
\ \forall t
\label{eq:gen_br_rp_ramp}\\
&0\le u_{g,t}\le K\overline R_g,\qquad
0\le d_{g,t}\le K\underline R_g,
\ \forall t
\label{eq:gen_br_rp_prod}\\
&q_{g,t}+u_{g,t}\le\overline P_g,\qquad
q_{g,t}-d_{g,t}\ge\underline P_g,
\ \forall t.
\label{eq:gen_br_rp_head}
\end{align}

\subsection{Realized Profit and Lost Opportunity Cost}
Let $p_{g,t}^{*}$ denote the dispatch assigned by the ISO under a given market design. For the energy-only settlements $m\in\{\mathrm{LA},\mathrm{TLMP}\}$, realized profit is
\begin{align}
\Pi_g^{\mathrm{disp},m} = \sum_{t=1}^T \left( \pi_{g,t}^{m} p_{g,t}^{*,m} - c_g p_{g,t}^{*,m} \right)\Delta t. \label{eq:profit_energy_only}
\end{align}
Under RP-LMP, realized profit includes both energy and ramp-product revenues:
\begin{align}
\Pi_g^{\mathrm{disp},RP} = \sum_{t=1}^T \left[(\lambda_t^{RP}-c_g)p_{g,t}^{*,RP} + \nu_t^{u} r_{g,t}^{u} + \nu_t^{d} r_{g,t}^{d}\right]\Delta t. \label{eq:profit_rp}
\end{align}

The ex-post lost opportunity cost (LOC) is the difference between best-response profit and realized profit $(m \in \{\text{LA}, \text{TLMP}\})$:
\begin{align}
LOC_g^{RP} := Q_g^{RP} - \Pi_g^{\mathrm{disp},RP},
LOC_g^{m} := Q_g(\pi_g^m) - \Pi_g^{\mathrm{disp},m} \label{eq:loc_def}
\end{align}

\begin{proposition}\label{prop:loc_decomp}
For any feasible dispatch $p$ that serves a fixed demand path $d$ with no load shedding under a common energy-price vector $\pi$,
\begin{align}
\sum_g LOC_g(p,\pi)
=
\underbrace{C(p)-C^\star}_{\text{dispatch inefficiency}}
+
\underbrace{\sum_g Q_g(\pi)-\pi^\top d + C^\star}_{\text{price support gap}}
\label{eq:loc_decomp}
\end{align}
where $C(p):=\sum_{g,t} c_g p_{g,t}\Delta t$, $\pi^\top d:=\sum_t\pi_t d_t\Delta t$, and $C^\star$ is the minimum feasible production cost for $d$. The identity follows from
\begin{align}
\sum_g LOC_g
&=\sum_g Q_g(\pi)-\left(\pi^\top d-C(p)\right) \nonumber\\
&=\left(C(p)-C^\star\right)
+\left(\sum_g Q_g(\pi)-\pi^\top d+C^\star\right). \nonumber
\end{align}
It applies directly to common energy-price settlements; RP product revenues are handled separately in \eqref{eq:profit_rp}.
\end{proposition}

Equation \ref{eq:loc_decomp} separates a dispatch effect from a price-support effect. LAED can reduce the first term by improving the implemented trajectory, but uniform current-interval prices cannot eliminate the second term when ramp constraints bind. TLMP removes this residual gap by embedding the generator-specific marginal value of ramping constraints into the settlement price.

\subsection{Generator-Level Explanatory Features}
To characterize cross-generator heterogeneity in LOC, we screen features constructed from the realized trajectories:
\begin{itemize}
    \item \textbf{Flexibility ratio:} $\mathrm{flex}_g=R_g/\overline P_g$.
    \item \textbf{Bind frequency:} the fraction of realized ramp movements that are nearly binding.
    \item \textbf{Flex-adjusted bind exposure:} $\mathrm{bind}_g/\mathrm{flex}_g$.
\end{itemize}
The multivariate analysis also controls for capacity, marginal cost, and realized utilization.

\section{Simulation Design}
\label{sec:setup}
\subsection{Rolling-Horizon Structure}
All dispatch problems use a deterministic, single-bus rolling-horizon simulation with 5-minute settlement intervals. At time $t$, the chosen dispatch model is solved over a look-ahead horizon, only the current dispatch decision is implemented, and the horizon is rolled forward. This setup uses a single-bus setting to isolate intertemporal ramping effects. Generator costs are linear, and the value of lost load is fixed at \mbox{\$3{,}500/MWh} \cite{Chen2023}.

\subsection{Test System and Implementation}
The reported results use two test systems. The first is a heterogeneous 10-generator MISO test system with varying costs, capacities, and ramp rates; Table~\ref{tab:gen_loc_example} reports its parameters together with the severe-case generator LOC results. The second is a controllable single-bus reduction of the RTS-GMLC system \cite{RTS-GMLC}, containing 93 generators.

\begin{table}[!t]
\caption{10-Generator system parameters and severe-case LOC. Cost is in dollars/MWh; capacity and 5-min ramp limit are in MW; LOC is in dollars. $\Delta LOC:=LOC_g^{RP}-LOC_g^{LA}$.}
\label{tab:gen_loc_example}
\centering
\small
\setlength{\tabcolsep}{2.3pt}
\begin{tabular}{crrrrrrr}
\toprule
Gen. & Cost & Cap. & Ramp & RP LOC & LA LOC & $\Delta LOC$ & TLMP LOC\\
\midrule
1  & 185.0 & 22  & 7.5  & 0.0    & 0.0             & 0.0 & 0\\
2  & 30.0  & 170 & 17.5 & 663.2  & \textbf{77.9}   & 585.4 & 0\\
3  & 55.0  & 85  & 7.5  & \textbf{52.5} & 101.6    & -49.1 & 0\\
4  & 15.0  & 230 & 7.5  & 42.8   & \textbf{0.0}    & 42.8 & 0\\
5  & 20.0  & 613 & 37.5 & 0.0    & 0.0             & 0.0 & 0\\
6  & 19.5  & 686 & 10.0 & 3487.3 & \textbf{1984.1} & 1503.3 & 0\\
7  & 48.0  & 45  & 10.0 & 47.0   & \textbf{8.9}    & 38.1 & 0\\
8  & 60.0  & 50  & 5.0  & \textbf{0.0} & 5.7      & -5.7 & 0\\
9  & 57.0  & 260 & 5.0  & \textbf{10.4} & 27.1    & -16.7 & 0\\
10 & 50.0  & 400 & 12.5 & 4129.8 & \textbf{278.0}  & 3851.8 & 0\\
\midrule
\multicolumn{4}{r}{Total LOC} & 8433.1 & \textbf{2483.1} & 5950.0 & 0\\
\bottomrule
\end{tabular}
\end{table}

We study three uniform ramp-scaling regimes for the 10-generator system: 0.8, 0.2, and 0.1 of the base ramp capability. The demand trajectory is held fixed while ramp capability is scaled, isolating ramp scarcity. 

Figure~\ref{fig:dispatch_compare} and Table~\ref{tab:gen_loc_example} use the deterministic 10-generator 0.1 ramp factor scenario with a 1-hour look-ahead horizon. The load trajectory is the August 2032 projection from MISO, scaled to a 1{,}100 MW reference capacity over the course of a whole day. The models are implemented as linear programs in Pyomo and solved with Gurobi.

The featured 10-generator system uses $U_t^u=40$ MW as a proxy for MISO's upward uncertainty adder, $U_t^d=0$, and $\kappa^u=\kappa^d=\$65/\text{MWh}$ as in \cite{Chen2023}. The RTS-GMLC system aggregates an 8-hour high-ramp window from 2020-06-11, scales peak load to 95\% of controllable capacity, and uses ramp factor 0.6.

Separate robustness checks use 28 full-day RTS-GMLC net-load profiles (seven January, April, July, and October days), scaled to 1,100 MW mean for the 10-generator system. This grid varies ramp multipliers (0.1, 0.2, 0.4, 0.8), upward-adder fractions $\beta \in \{0, 0.08, 0.16, 0.24\}$ with $U^u_t=\beta \bar{d}$, LAED horizons $\{1, 3, 7, 13\}$, and error scales $\{0,1,3,5\}$\%. Forecasts use Gaussian AR(1) perturbations ($\phi=0.9$), shared by both designs and retained as windows roll; scale grows with the square root of lead time, capped at four intervals.

\section{Results}
\label{sec:results}
\subsection{Cross-Regime Summary}
On the featured day, tightening the ramp factor from 0.8 to 0.2 and 0.1 increases the share of generator ramp  constraint bindings from 0\% to 0.40\% and 3.59\%, respectively. Correspondingly, RP-LMP LOC increases from \$0 to \$187.40 and \$8,433.10, while LA-LMP LOC increases from \$0 to \$7.50 and \$2,483.10. At the most restrictive ramp factor of 0.1, 99 ramp transitions reach their limits with no load shedding. The upward ramp capability (RCUP) price is positive in 10.5\% of intervals and reaches \$40/MWh, while the largest difference between TLMP and the uniform energy price is \$44/MWh.

In this severe case, LA-LMP reduces total LOC by 70.6\%, from \$8,433.10 under RP-LMP to \$2,483.10, and LOC is completely eliminated by TLMP as shown in Table~\ref{tab:gen_loc_example}. Thus, look-ahead pricing substantially mitigates aggregate LOC but does not eliminate it, because a single uniform energy price cannot fully compensate every generator for the opportunity costs induced by binding ramp constraints \cite{Guo2021}. By contrast, when ramp constraints do not bind at a ramp factor of 0.8, both RP-LMP and LA-LMP yield zero LOC. The emergence of LOC as ramp capability tightens therefore shows that the burden is driven by binding ramp constraints rather than by the presence of a ramp product alone.

Because the two designs change both dispatch and settlement, we perform two comparisons on an independent full-day load profile. Holding RP dispatch fixed, adding ramp-capability revenue reduces LOC from \$10{,}521 to \$5{,}417. Holding LA-LMP settlement fixed and extending the horizon from 1 to 13 intervals reduces LOC from \$5{,}572 to \$280. This illustrates the simultaneous impact of both components of the LOC decomposition in Proposition \ref{prop:loc_decomp}.

\subsection{Which Generators Bear the LOC Burden?}
Unless noted otherwise, results below refer to the 10-generator ramp-stressed scenario. Table~\ref{tab:gen_loc_example} shows two basic results: both RP-LMP and LA-LMP yield positive LOC for some generators; and neither RP-LMP nor LA-LMP produces lower LOC for every generator. 

Although total LOC is lower under LA-LMP, not every generator benefits. LA-LMP lowers LOC for five generators, RP-LMP lowers it for three, and two are tied. Generator 10 is the clearest example: under RP-LMP its realized profit is \mbox{-\$1{,}961.10}, even though it could have earned a positive profit by choosing its own best feasible dispatch at the same prices.

Generators 10, 6, and 2 have the largest RP-LMP LOC, and generators 10 and 6 account for 90.3\% of the total. LA-LMP lowers LOC for generators 10, 6, and 2 by \$3{,}851.80, \$1{,}503.30, and \$585.40, respectively. For the three generators with higher LOC under LA-LMP, the increases are only \$49.1, \$5.7, and \$16.7. Thus aggregate LOC reduction under LA-LMP comes primarily from relieving the generators that bear most of the RP-LMP burden, while this offsetting increases LOC for other generators by a smaller amount.

The larger RTS-GMLC system shows similar LOC concentration. Positive LOC occurs for 55 of 93 generators under RP-LMP and only 26 under LA-LMP. Only 11 generators have higher LOC under LA-LMP, by an average of \$3.30. In both designs, the largest burdens fall on a small group of gas combined-cycle units.

\subsection{What Drives LOC Across Generators?}
\begin{figure}[!t]
    \centering
    \includegraphics[width=\columnwidth,trim=0 8 0 8,clip]{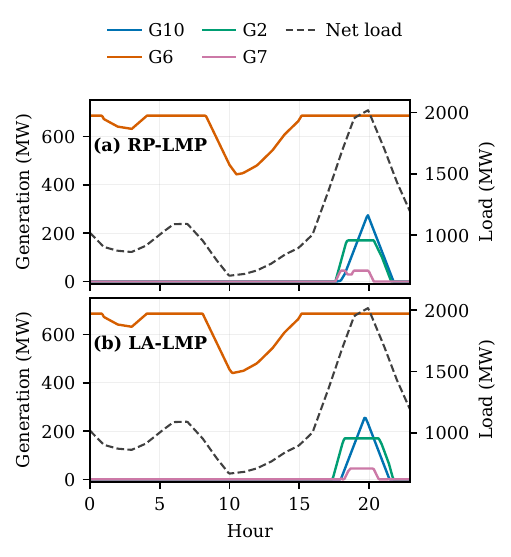}
    \caption{Full-day dispatches for 4 generators in the 10-generator ramp-constrained case under RP-LMP (top) and LA-LMP (bottom).}
    \label{fig:dispatch_compare}
\end{figure}

The source of LAED's aggregate LOC reduction becomes clear in the generator-level dispatch patterns in Figure~\ref{fig:dispatch_compare}: LAED moves frequently constrained generators earlier, reducing repeated ramp binding. These dispatch patterns align with generator characteristics. Generators 10, 6, and 2 have flexibility ratios of 0.031, 0.015, and 0.103, and their ramps bind in 14.3\%, 7.8\%, and 3.4\% of intervals. Generator 10 reaches its upward ramp limit in 6.9\% of intervals in which generating more power would be profitable; its average price-cost margin in those intervals is \mbox{\$1.82/MWh}. Generator 6 instead has very low flexibility.

In the most ramp-constrained 10-generator case, flexibility ratio alone explains little of the variation in LOC ($R^2=0.158$ for RP-LMP and 0.113 for LA-LMP). For RP-LMP, ramp-bind frequency gives $R^2=0.823$, which rises to 0.849 when bind frequency is divided by flexibility. For LA-LMP, the corresponding values are 0.256 and 0.697. How often a generator's ramp binds is therefore a stronger indicator of LOC than low flexibility alone, although low flexibility makes these events more costly. These generators also receive the largest LA-LMP relief.

\begin{figure*}[!t]
    \centering
    \includegraphics[width=\textwidth]{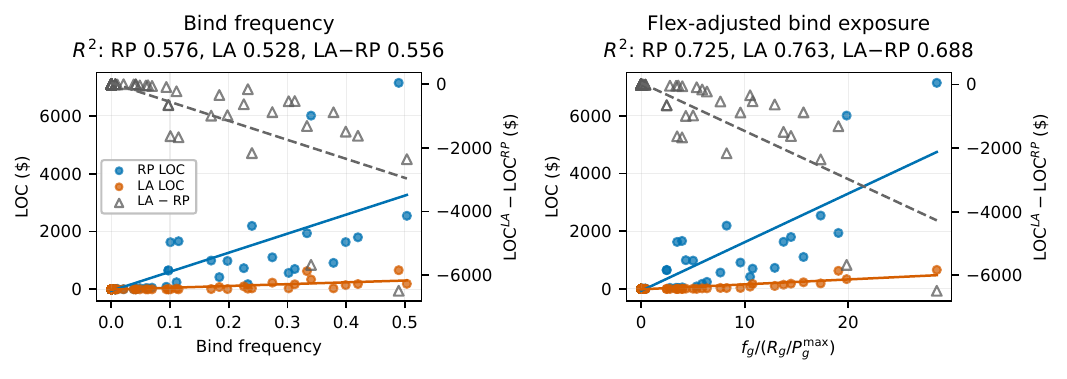}
    \caption{Generator-level LOC under RP-LMP and LA-LMP in the RTS-GMLC system, with $LOC^{LA}-LOC^{RP}$ on the right axis. The left panel uses ramp-bind frequency, and the right divides this by flexibility ratio. RP-LMP LOC and LA-LMP reduction are largest for generators whose ramps bind often.}
    \label{fig:rts_bivariate}
\end{figure*}

Figure~\ref{fig:rts_bivariate} confirms this pattern for one RTS-GMLC day. Ramp-bind frequency alone gives $R^2=0.576$ for RP-LMP LOC and 0.556 for $\Delta LOC:=LOC^{RP}-LOC^{LA}$; dividing it by flexibility raises these values to 0.725 and 0.688. We then estimate two generator-day regressions across both systems using cases without loadshed (1{,}256 observations from 103 generators). The outcomes, reported in the same order below, are $\log(1+\mathrm{RP\ LOC/MW})$ and the signed log of LA-LMP relief, $(\mathrm{RP}-\mathrm{LA})$ LOC per MW. The signed log preserves the difference direction and limits extreme values' influence. Each regression includes bind frequency, log flexibility, log capacity, marginal cost, utilization, each standardized within its test-system; p-values use generator-clustered standard errors to account for repeated observations. Bind frequency is positive in both regressions: 0.864 ($p<0.001$) and 0.819 ($p<0.001$). Log flexibility is negative: $-0.223$ ($p=0.017$) and $-0.219$ ($p=0.019$), as is log capacity: $-0.200$ ($p=0.037$) and $-0.204$ ($p=0.034$). Marginal cost ($p=0.076$ and 0.070) and utilization ($p=0.140$ and 0.157) are not significant at a 5\% significance level. The regressions have $R^2=0.812$ and 0.802.

\subsection{Validation and Forecast Error}

The larger RTS-GMLC system shows the same overall result. At ramp multiplier 0.6, $U_t^u=0$, so the ramp requirement contains only forecasted net-load change; both RP-LMP and LA-LMP serve all load. The RCUP price reaches \mbox{\$9.06/MWh}. Total LOC is $\$53{,}457$ under RP-LMP, $\$1{,}463$ under LA-LMP, and zero under TLMP, so the RP-LMP result is not caused by load shedding. 

Table~\ref{tab:robustness_summary} summarizes the multi-day validation and forecast-sensitivity checks. Under Gaussian AR(1) forecast errors, median $\Delta LOC$ changes from positive to negative between 1\% and 3\% forecast error. Thus, LA-LMP has lower median LOC at the smaller error level but higher median LOC at the larger level. LAED is more exposed to persistent forecast errors because its current dispatch is shaped by the full 13-interval forecast path, whose errors grow with lead time, whereas RP dispatch uses realized current net load and only the 10-minute forecast to set its ramp requirement. Because this comparison uses one stylized AR(1) process and different forecast horizons in a deterministic single-bus model, the crossover indicates sensitivity of this particular implementation, not necessarily that LA-LMP pricing is inherently less robust to forecast perturbations.

\begin{table}[!t]
\caption{LOC across multiple days. No shed gives fraction of cases without loadshed and $\Delta LOC > 0$ is the percent of intervals where LA has lower LOC. $\Delta LOC$ is in dollars; IQR is interquartile range}
\label{tab:robustness_summary}
\centering
\scriptsize
\setlength{\tabcolsep}{1.8pt}
\renewcommand{\arraystretch}{0.90}
\begin{tabular}{lrrrr}
\toprule
Case & No shed & $\Delta LOC > 0$ & Median $\Delta LOC$ & IQR $\Delta LOC$ \\
\midrule
\multicolumn{5}{l}{10-generator, perfect foresight, adder $\beta=0.16$} \\
Ramp 0.2 & 6/28 & 66.7\% & 149 & [25, 413] \\
Ramp 0.4 & 14/28 & 71.4\% & 23 & [1, 79] \\
Ramp 0.8 & 20/28 & 15.0\% & 0 & [0, 0] \\
\midrule
\multicolumn{5}{l}{10-generator, AR(1), ramp 0.2, adder $\beta=0.08$} \\
0\% error & 6/28 & 66.7\% & 149 & [25, 413] \\
1\% error & 18/84 & 50.0\% & 50 & [$-33$, 563] \\
3\% error & 19/84 & 10.5\% & $-1{,}162$ & [$-2{,}829$, $-426$] \\
5\% error & 19/84 & 0.0\% & $-10{,}603$ & [$-18{,}185$, $-8{,}855$] \\
\midrule
\multicolumn{5}{l}{RTS-GMLC, perfect foresight, adder $\beta=0.16$} \\
Ramp 0.2 & 12/12 & 100.0\% & 62{,}199 & [52{,}313, 78{,}329] \\
Ramp 0.4 & 12/12 & 100.0\% & 31{,}981 & [18{,}144, 46{,}441] \\
Ramp 0.8 & 12/12 & 100.0\% & 8{,}330 & [3{,}754, 13{,}715] \\
\midrule
\multicolumn{5}{l}{RTS-GMLC, AR(1), ramp 0.6, adder $\beta=0.16$} \\
0\% error & 12/12 & 100.0\% & 15{,}448 & [5{,}500, 23{,}073] \\
1\% error & 36/36 & 88.9\% & 12{,}164 & [1{,}465, 19{,}122] \\
3\% error & 36/36 & 30.6\% & $-10{,}797$ & [$-20{,}659$, 1{,}280] \\
5\% error & 36/36 & 5.6\% & $-48{,}164$ & [$-73{,}345$, $-24{,}478$] \\
\bottomrule
\end{tabular}
\renewcommand{\arraystretch}{1}
\end{table}

To isolate uncertainty in the ramp requirement from forecast error, $\beta$ scales the upward adder $U_t^u=\beta\bar d$. Across the reported ramp factors, qualifying perfect-foresight cases at every tested $\beta$ satisfy aggregate \mbox{$LOC^{LA}\le LOC^{RP}$}; the remainder are ties. For RTS-GMLC, median LOC relief declines from \$62{,}199 at ramp factor 0.2 to \$31{,}981 at 0.4 and \$8{,}330 at 0.8, with LA-LMP lower on all 12 days. Across ten RTS-GMLC days, median LA-LMP LOC falls from \$50{,}634 with one interval to \$23{,}876, \$11{,}812, and \$5{,}405 with 3, 7, and 13 intervals, respectively. RP-LMP does not change because it has no look-ahead window.

\section{Conclusion}
\label{sec:conclusion}
This paper compares RP-LMP, LA-LMP, and TLMP in a common rolling-horizon framework under ramp stress. In the deterministic cases, RP-LMP leaves larger aggregate ex-post LOC than LA-LMP, concentrated on generators with repeated ramp-binding exposure. LA-LMP relieves those same units but, under a uniform current-interval price, does not eliminate residual LOC; TLMP eliminates it. Multi-day perfect-foresight checks preserve the aggregate ordering in the larger system, while forecast-error tests show that it is not universal. Because the RP representation is stylized and omits network, commitment, and detailed ISO demand-curve rules, these results do not establish generic superiority of one market implementation. They instead show why ramp-product designs should be evaluated both for procured capability and for compensation of generators that repeatedly absorb intertemporal scarcity.

\section*{Disclaimer}
The views expressed in this paper are the opinion of the authors and do not reflect the views of PJM Interconnection, L.L.C. or its Board of Managers of which Le Xie is a member.

\bibliographystyle{IEEEtran}
\bibliography{references}

@ARTICLE{Guo2021,
  author={Guo, Ye and Chen, Cong and Tong, Lang},
  journal={IEEE Transactions on Power Systems}, 
  title={Pricing Multi-Interval Dispatch Under Uncertainty Part I: Dispatch-Following Incentives}, 
  year={2021},
  volume={36},
  number={5},
  pages={3865-3877},
  doi={10.1109/TPWRS.2021.3055730}}

@ARTICLE{Zhao2020,
  author={Zhao, Jinye and Zheng, Tongxin and Litvinov, Eugene},
  journal={IEEE Transactions on Power Systems}, 
  title={A Multi-Period Market Design for Markets With Intertemporal Constraints}, 
  year={2020},
  volume={35},
  number={4},
  pages={3015-3025},
  doi={10.1109/TPWRS.2019.2963022}}

@ARTICLE{Hua2019,
  author={Hua, Bowen and Schiro, Dane A. and Zheng, Tongxin and Baldick, Ross and Litvinov, Eugene},
  journal={IEEE Transactions on Power Systems}, 
  title={Pricing in Multi-Interval Real-Time Markets}, 
  year={2019},
  volume={34},
  number={4},
  pages={2696-2705},
  doi={10.1109/TPWRS.2019.2891541}}

@ARTICLE{Scott2019,
  author={Scott, Paul and Thiébaux, Sylvie},
  journal={IEEE Transactions on Smart Grid}, 
  title={Identification of Manipulation in Receding Horizon Electricity Markets}, 
  year={2019},
  volume={10},
  number={1},
  pages={1046-1057},
  doi={10.1109/TSG.2017.2758394}}

@techreport{mickey2015mirtm,
  author       = {Mickey, J.},
  title        = {Multi-Interval Real-Time Market Overview},
  institution  = {ERCOT},
  year         = {2015},
  month        = {October},
  note         = {Board of Directors Meeting Presentation},
  url          = {https://www.ercot.com/files/docs/2015/10/06/5_Multi_Interval_Real_Time_Market_Overview.pdf}
}

@ARTICLE{ela2016,
  author={Ela, Erik and O'Malley, Mark},
  journal={IEEE Transactions on Power Systems}, 
  title={Scheduling and Pricing for Expected Ramp Capability in Real-Time Power Markets}, 
  year={2016},
  volume={31},
  number={3},
  pages={1681-1691},
  doi={10.1109/TPWRS.2015.2461535}}

@ARTICLE{Wang2017_review,
  author={Wang, Qin and Hodge, Bri-Mathias},
  journal={IEEE Transactions on Industrial Informatics}, 
  title={Enhancing Power System Operational Flexibility With Flexible Ramping Products: A Review}, 
  year={2017},
  volume={13},
  number={4},
  pages={1652-1664},
  doi={10.1109/TII.2016.2637879}}

@ARTICLE{CWang2017,
  author={Wang, Congcong and Bao-Sen Luh, Peter and Navid, Nivad},
  journal={IEEE Transactions on Power Systems}, 
  title={Ramp Requirement Design for Reliable and Efficient Integration of Renewable Energy}, 
  year={2017},
  volume={32},
  number={1},
  pages={562-571},
  doi={10.1109/TPWRS.2016.2555855}}

@ARTICLE{wu2016,
  author={Wu, Chenye and Hug, Gabriela and Kar, Soummya},
  journal={IEEE Transactions on Power Systems}, 
  title={Risk-Limiting Economic Dispatch for Electricity Markets With Flexible Ramping Products}, 
  year={2016},
  volume={31},
  number={3},
  pages={1990-2003},
  doi={10.1109/TPWRS.2015.2460748}}

@ARTICLE{Chen2017,
  author={Chen, Runze and Wang, Jianhui and Botterud, Audun and Sun, Hongbin},
  journal={IEEE Transactions on Power Systems}, 
  title={Wind Power Providing Flexible Ramp Product}, 
  year={2017},
  volume={32},
  number={3},
  pages={2049-2061},
  doi={10.1109/TPWRS.2016.2603225}}

@ARTICLE{Chen2023,
  author={Chen, Yonghong},
  journal={IEEE Transactions on Power Systems}, 
  title={Addressing Uncertainties Through Improved Reserve Product Design}, 
  year={2023},
  volume={38},
  number={4},
  pages={3911-3923},
  doi={10.1109/TPWRS.2022.3200697}}

@misc{hogan2020,
author      = {Hogan, William W.},
title       = {Electricity Market Design: Multi-interval Pricing Models},
year        = {2020},
month       = jun,
url         = {https://whogan.scholars.harvard.edu/sites/g/files/omnuum4216/files/whogan/files/hogan_hepg_multi_period_062220.pdf}}

@article{Cho2023,
author = {Cho, Jehum and Papavasiliou, Anthony},
title = {Pricing Under Uncertainty in Multi-Interval Real-Time Markets},
journal = {Operations Research},
volume = {71},
number = {6},
pages = {1928-1942},
year = {2023},
doi = {10.1287/opre.2022.2314},
URL = {https://doi.org/10.1287/opre.2022.2314},
eprint = {https://doi.org/10.1287/opre.2022.2314}
}

@misc{Zhang2026_PSCC,
      title={Comparative Assessment of Look-Ahead Economic Dispatch and Ramp Products for Grid Flexibility}, 
      author={Qian Zhang and Le Xie and Long Zhao and Congcong Wang},
      year={2026},
      eprint={2601.22120},
      archivePrefix={arXiv},
      primaryClass={eess.SY},
      url={https://arxiv.org/abs/2601.22120}, 
}

@ARTICLE{RTS-GMLC,
  author={Barrows, Clayton and Bloom, Aaron and Ehlen, Ali and Ikäheimo, Jussi and Jorgenson, Jennie and Krishnamurthy, Dheepak and Lau, Jessica and McBennett, Brendan and O’Connell, Matthew and Preston, Eugene and Staid, Andrea and Stephen, Gord and Watson, Jean-Paul},
  journal={IEEE Transactions on Power Systems}, 
  title={The IEEE Reliability Test System: A Proposed 2019 Update}, 
  year={2020},
  volume={35},
  number={1},
  pages={119-127},
  doi={10.1109/TPWRS.2019.2925557}}

@techreport{MISO2026URC,
  author={{MISO Market Subcommittee}},
  title={Evaluation of Updates to the Up Ramp Capability Demand Curve},
  institution={Midcontinent Independent System Operator},
  year={2026},
  month=apr,
  note={Item MSC-2026-1}
}

\end{document}